\documentclass[sigconf]{acmart}
\usepackage{url}
\usepackage{hyperref}  
\usepackage{tabularx}  
\usepackage{booktabs}
\usepackage{hyperxmp}  

\usepackage{acmart-taps}
\AtBeginDocument{%
  \providecommand\BibTeX{{%
    \normalfont B\kern-0.5em{\scshape i\kern-0.25em b}\kern-0.8em\TeX}}}

\copyrightyear{2025}
\acmYear{2025}
\setcopyright{rightsretained}
\acmConference[CHI EA '25]{Extended Abstracts of the CHI Conference on Human Factors in Computing Systems}{April 26-May 1, 2025}{Yokohama, Japan}
\acmBooktitle{Extended Abstracts of the CHI Conference on Human Factors in Computing Systems (CHI EA '25), April 26-May 1, 2025, Yokohama, Japan}
\acmDOI{10.1145/3706599.3720130}
\acmISBN{979-8-4007-1395-8/2025/04}

\begin{document}

\title[Virtual Co-presenter]{Virtual Co-presenter: Connecting Deaf and Hard-of-hearing Livestreamers and Hearing audience in E-commerce Livestreaming}


\author{Yuehan Qiao}
\email{yhqiao@umd.edu}
\affiliation{%
  \institution{Tsinghua University}
  \streetaddress{30 Shuangqing Rd}
  \city{Beijing}
  \country{China}
}

\author{Zhihao Yao}
\email{yaozh_h@outlook.com}
\affiliation{%
  \institution{Tsinghua University}
  \streetaddress{30 Shuangqing Rd}
  \city{Beijing}
  \country{China}
}

\author{Meiyu Hu}
\email{my2424huuu@gmail.com}
\affiliation{%
  \institution{Tongji University}
  \city{Shanghai}
  \country{China}}

\author{Qianyao Xu}
\email{xuqianyao@yeah.net}
\affiliation{%
  \institution{Tsinghua University}
  \streetaddress{30 Shuangqing Rd}
  \city{Beijing}
  \country{China}
}
\authornote{Corresponding Author.}

\begin{abstract}

  Deaf and Hard-of-Hearing (DHH) individuals are increasingly participating as livestreamers in China's e-commerce livestreaming industry but face obstacles that limit the scope and diversity of their audience. Our paper examines these challenges and explores a potential solution for connecting the hearing audience to sign language (SL) livestreaming teams with DHH members in e-commerce livestreaming. We interviewed four SL livestreaming team members and 15 hearing audience members to identify information and emotional communication challenges that discourage the hearing audience from continuing to watch SL livestreaming. Based on these findings, we developed a virtual co-presenter demo, which targets SL livestreaming teams with DHH members as users, through a design workshop with six designers, incorporating voice broadcasting with animations. Follow-up evaluations with previous participants provided positive feedback on the virtual co-presenter’s potential to address these challenges. We summarize design suggestions on its functionality and interaction design for further refinement to assist SL livestreaming teams with DHH members in reaching a broader hearing audience.

\end{abstract}

\ccsdesc[500]{Human-centered computing~Empirical studies in accessibility}

\keywords{Deaf and hard-of-hearing (DHH), livestreaming, virtual human, accessibility}



\maketitle

\section{Introduction}

Livestreaming has become a dominant form of social media entertainment worldwide, creating employment opportunities across various sectors\cite{10.1145/3491102.3517634,10.1145/3290605.3300459}, including for individuals with disabilities\cite{johnson2019inclusion,jun2021exploring,rong2022feels,anderson2022gamer}. In e-commerce livestreaming, which relies heavily on verbal communication and real-time interaction\cite{doi:10.1080/10447318.2022.2076773}, Deaf and Hard of Hearing (DHH) livestreamers face unique challenges\cite{cao2023sparkling}. They primarily communicate through Sign Language (SL)\cite{yanxiaojing2021taobao}, which inherently limits their audience to the DHH community, reducing the hearing audience reach and earning potential of DHH livestreamers.

Prior research has explored SL conversion technologies to facilitate communication between DHH and hearing individuals\cite{ahmed2021real,ahmed2018review,camgoz2018neural,camgoz2020sign}. However, e-commerce livestreaming requires not only accurate information transmission but also engaging presentation skills, which SL conversion technologies struggle to provide. Recently, virtual livestreamers have gained popularity among younger audiences due to their enhanced expressiveness \cite{doi:10.1177/0022242921996646,liew2018exploring,GAO2023103356}, yet little research has examined how virtual human technology could support DHH livestreamers. Inspired by this, we explore the potential of virtual human technology in assisting DHH livestreamers in SL e-commerce livestreaming.

To comprehensively examine this issue, our research adopts a bilateral perspective, investigating communication barriers in SL e-commerce livestreaming from both the perspectives of SL livestreaming teams with DHH members and hearing audience. Specifically, we address two key research questions: (1) What are the primary challenges SL livestreaming teams with DHH members face in connecting with hearing audience? (2) What factors deter hearing audience from sustained engagement in SL livestreaming? Through semi-structured interviews with four SL livestreaming team members and 15 hearing audiences, followed by inductive coding and thematic analysis, we identified communication gaps in information conveyance and emotional resonance between these two groups, informing our assistive system design.

Building on these findings, we conducted a professional design workshop with six designers to explore potential solutions. Through collaborative efforts, we developed a virtual co-presenter system which targets SL livestreaming teams with DHH members as users and is designed to bridge communication gaps in livestreaming to hearing audiences. To evaluate the design, we conducted follow-up assessments with both SL livestreaming team members and hearing audience, gathering feedback on system functionality and interaction design. These evaluations provided insights into the system’s practical applications and future development directions in real livestreaming scenarios.

Our research offers two main contributions: (1) An in-depth qualitative investigation of the challenges faced by SL livestreaming teams with DHH members and hearing audience, examining their perspectives on communication barriers in e-commerce livestreaming; (2) The design and preliminary evaluation of a virtual co-presenter demo aimed at supporting SL livestreaming teams with DHH members in reaching a broader hearing audience.

\begin{figure*}[h!]
    \centering 
    \includegraphics[width=\linewidth]{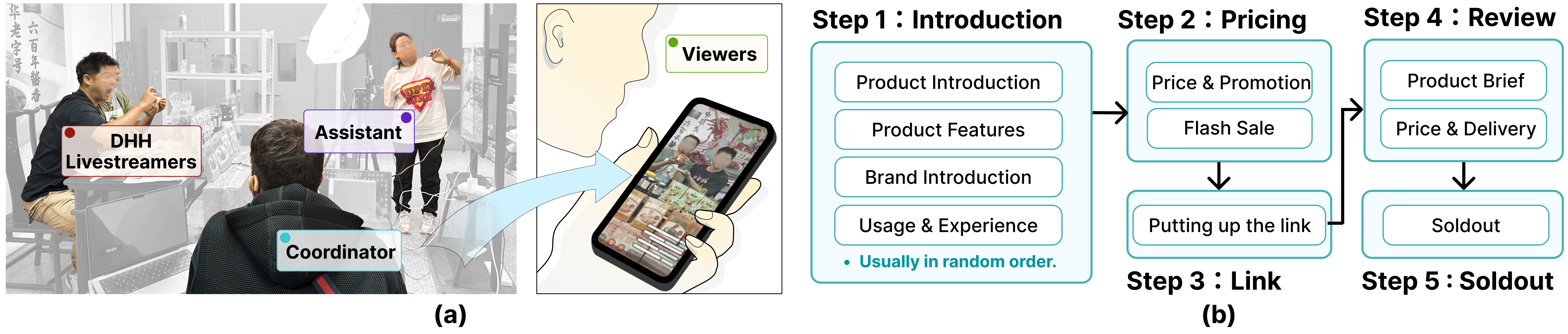}
    \caption{The setup of an e-commerce livestreaming. (a) the behind-the-scenes team composition of the SL e-commerce livestreaming viewed on mobile phones; (b) the common process in e-commerce livestreaming.}
    \label{fig:comp&process}
\end{figure*}

\section{Sign language E-commerce livestreaming}

Sign language e-commerce livestreaming is a specialized form of e-commerce livestreaming where DHH livestreamers promote products or services using SL and textual prompts while interacting with audience in real time\cite{yanxiaojing2021taobao,cao2023sparkling}. A typical SL livestreaming team, primarily composed of DHH individuals, consists of three key roles: livestreamer, coordinator, and assistant (Fig. \ref{fig:comp&process}a). Each team includes at least one DHH livestreamer, while others may be DHH or hearing individuals proficient in SL.

The preparation and livestreaming process follow a fixed and structured workflow. Before the livestream, the team collaborates with brands to finalize product details, promotional strategies, and a detailed script that the team will follow in livestreaming. During the livestream, DHH livestreamers adhere to a defined sequence (Fig. \ref{fig:comp&process}b). It includes introducing products, explaining pricing and promotions, guiding audience through purchasing, and providing real-time sales updates while engaging with the audience. Meanwhile, the coordinator manages the livestreaming process behind the stage and interacts with audience via live comments, while the assistant handles teleprompting tasks.

\section{Related Works}

\subsection{DHH content creators on social media}

With the rise of social media, DHH individuals have gained opportunities to participate as content creators\cite{schafer2023social,infoguides_lead_301,10.1145/3491102.3517574}, but their engagement remains limited. Their contributions primarily involve asynchronous content, such as uploading edited videos. In these videos, DHH creators use captions, AI-generated voices, and other accessibility tools to communicate with a broader audience\cite{10.1145/3544548.3581286,10.1145/3613904.3642413}. Beyond asynchronous content, some DHH individuals have begun participating in livestreaming as livestreamers\cite{cao2023sparkling}.

However, livestreaming, which relies on real-time interaction and dynamic expressiveness\cite{liu2025m,lu2018you,doi:10.1080/10447318.2022.2076773}, poses challenges for DHH content creators\cite{cao2023sparkling}. Methods commonly used for asynchronous content creation are often insufficient for livestreaming, particularly in e-commerce livestreaming, where strong real-time engagement is crucial\cite{wang2022live,chen2020study,joo2023perceived,wang2024live}.

Despite the high demands for real-time interaction and expressiveness in e-commerce livestreaming, its pre-structured and relatively fixed interaction processes\cite{kang2021dynamic,liu2025m} still offer design opportunities for enhancing communication in SL livestreaming teams with DHH members. Therefore, our study focuses on this unique scenario, exploring how to better support these teams by leveraging the structured nature of e-commerce livestreaming to improve engagement with a wider scope audience.

\subsection{Virtual human supporting DHH communication}

The rapid development and widespread adoption of virtual human (VH)
let it emerge as a possible solution for providing accessible information for DHH individuals by generating SL animations\cite{10.1145/3025453.3025642,10.1145/2556288.2557048,do2010towards,llamazares2021sign}. For instance, the Beijing Winter Olympics AI signer launched in 2022\cite{CDPF}, and some applications include AI SL streamers like Nai Nai introduced by iQiyi in 2018 \cite{CNDCM}, as well as the SignPal Kit in 2021 \cite{HUAWEI}. These applications developed by leading technology companies primarily focus on helping DHH individuals as consumers of audio-visual media with significantly high costs. However, limited research has focused on the use of VH to help DHH individuals as creators express their content. Our work is inspired by the VH applications and tries to bridge this gap in SL e-commerce livestreaming. We use VH to bridge SL livestreaming team with DHHs and hearing audiences to support DHH's livestreaming career.

\section{Interviews for Understanding Challenges}

In this section, we aimed to understand the communication barriers between SL livestreaming teams with DHH members and hearing audience in SL e-commerce livestreaming to inform the following design workshop.

\subsection{Methods}
Following approval from the university’s Institutional Review Board, we recruited four SL livestream team members (DL1-DL4) by sending recruitment messages on Kuaishou livestream app and snowball sampling in local special education institutions. Three are DHH livestreamers and one is a hearing individual working in an SL livestream team with DHH livestreamers. We also recruited 15 hearing audiences (HV1-HV15), aged 18-40 (M=25.33, SD=4.35; 9 males and 6 females), who were familiar with e-commerce livestreaming and watched livestreams at least twice a week for more than 30 minutes. Full demographic details are provided in Table \ref{tab:DHHs1} and Appendix A Table \ref{tab:hearingviewer}.

Semi-structured interviews were conducted with participants' consent. In interviews with DLs, we explored their livestreaming experiences and the challenges they faced in expanding the scope and diversity of their audience. In interviews with HVs, we showed two SL livestream recordings and asked about their viewing experience. All interviews were conducted via WeChat and Tencent Meeting. Since we did not involve SL interpreters, interviews with DL1 and DL3 were conducted through text messages on WeChat, while DL4, who used hearing aids, and all hearing participants completed the interviews via online meetings. The interviews with DLs lasted between 1 hour to 3 hours and the interviews with HVs lasted between 18 minutes to 33 minutes excluding the time spent watching SL livestream recordings. All text messages were documented, and all meetings were audio-recorded and transcribed verbatim in Chinese.

After finishing the interviews, we analyzed the transcripts using iterative inductive coding \cite{huberman2014qualitative} and reflexive thematic analysis \cite{Virginia2006Usingthematic}. First, we reviewed the transcripts to familiarize ourselves with the data. The first and second authors performed iterative open coding for multiple rounds to identify themes. All authors then met to discuss and finalize the codebook. After open coding, the first and second authors followed the codebook to complete the coding and translated the selected quotes. The coding results and interpretations were double-checked by the other two authors.

\begin{table*}
\caption{Summary of DHH livestream team members in Study 1.}
\label{tab:DHHs1}
\begin{tabular}{cclccll}
\toprule
\textbf{ID} &
  \textbf{Gender} &
  \multicolumn{1}{c}{\textbf{Platforms}} &
  \textbf{\begin{tabular}[c]{@{}c@{}}History\\ (years)\end{tabular}} &
  \textbf{\begin{tabular}[c]{@{}c@{}}Frequency\\ (times/month)\end{tabular}} &
  \multicolumn{1}{c}{\textbf{Streaming Topics}} &
  \multicolumn{1}{c}{\textbf{Hearing Status}} \\ \hline
DL1 &
  F &
  \begin{tabular}[c]{@{}l@{}}Douyin \& \\ Kuaishou\end{tabular} &
  1 &
  4 &
  Electronic products &
  \begin{tabular}[c]{@{}l@{}}Hard of hearing, \\ Acquired hearing loss\end{tabular} \\ \hline
DL2 &
  M &
  Kuaishou &
  3 &
  \textless{}10 &
  \begin{tabular}[c]{@{}l@{}}Daily Necessities, \\ Food, Cosmetics, \\ Electronic products\end{tabular} &
  Hearing \\ \hline
DL3 &
  F &
  Kuaishou &
  3 &
  \textless{}10 &
  \begin{tabular}[c]{@{}l@{}}Daily Necessities, \\ Food, Cosmetics, \\ Electronic products\end{tabular} &
  \begin{tabular}[c]{@{}l@{}}Deaf, \\ Acquired hearing loss\end{tabular} \\ \hline
DL4 &
  M &
  \begin{tabular}[c]{@{}l@{}}Douyin \& \\ Kuaishou\end{tabular} &
  2 &
  8 &
  \begin{tabular}[c]{@{}l@{}}Daily Necessities, \\ Food\end{tabular} &
  \begin{tabular}[c]{@{}l@{}}Hard of hearing, \\ Acquired hearing loss\end{tabular} \\ \bottomrule
\end{tabular}
\end{table*}

\subsection{Challenges of reaching hearing audience in sign language livestreaming}

In this phase of the interview, we revealed that the absence of hearing audience in SL livestreaming by team with DHHs is due to hearing audience not establishing the intention to continue watching. The reasons for this phenomenon are the challenges in conveying and comprehending product information and the challenges in emotional expression and resonance.

Our interviews revealed that the absence of hearing audience in SL livestreaming by teams with DHH members is primarily due to hearing audience's lack of intention to continue watching. This issue stems from challenges in presenting and receiving product information, as well as challenges in emotional expression and resonance.

\subsubsection{Lack of intention to continue watching}

Hearing audience typically enter SL livestreams conducted by DHH livestreamers out of curiosity or compassion, but these motivations are insufficient to sustain long-term engagement. HV5 said, "\textit{Because of curiosity, I might click to take a look, but just for a couple of minutes.}" HV3 noted his willingness to support, "\textit{I would support as much as I can because the livestreamer is Deaf.}" However, HV6 pointed out that such purchases are often one-time acts of goodwill. She stated, "\textit{After expressing my goodwill, I may not watch it again.}" The following two sections outline the key reasons why hearing audience lose interest quickly.

\subsubsection{Challenges in Presenting and Receiving Product Information}

Hearing audience struggle to access product information presented by DHH livestreamers in SL e-commerce livestreaming due to both visual overload and ineffective auditory cues.

\textbf{Difficulty in obtaining SL information.}
Both DHH livestreamers and hearing audiences identified SL as a major communication barrier. DL2 stated, "\textit{Hearing audience might find it meaningless and leave immediately because they can't hear any speech or understand SL.}" HV2 added, "\textit{I don’t expect to get product information from SL livestreaming, and if there are hearing livestreamers selling the same products, I probably won’t buy from DHH livestreamers.}" Due to the challenges of SL translation, DL2's team abandoned plans to expand their viewer pool to hearing audience. DL4, facing the same issue, explained why collaboration with interpreters was not feasible. He said, "\textit{There are too few SL interpreters, and they are too expensive to hire.}"

\textbf{Confusing and overwhelming visual information.}
DHH livestreamers use text prompts to provide product details for audience with varying SL proficiency. DL2 described their approach, "\textit{We usually use a whiteboard or paper to display product qualifications and features.}" While HV8, HV11, and HV12 found written product details helpful, others felt the excessive text without clear explanation made comprehension difficult. HV6 said, "\textit{It’s too 'noisy' to my eyes. They just wrote down lots of numbers, and I don't know which one is the product's price without verbal explanation.}" HV1 echoed this concern, "\textit{It’s overwhelming to watch SL gestures, look at the product, and read the text at the same time.}" HV13 noted that she missed critical moments, such as livestreamers putting up the purchasing link, due to visual overload.

\textbf{Disruptive auditory interferences.} 
Although DHH livestreamers do not speak during SL livestreaming, various background noises are still present, such as sounds of body collisions during the signing, non-semantic vocalizations, and ambient sounds. These sounds were often unnoticed by DHH livestreamers, but were distracting to hearing audience. HV11 reported, "\textit{There are lots of noises, and they are very loud to me.}" HV1 added, "\textit{Not only is the sound meaningless, but it also interferes with my viewing experience. Since I know the livestreamer won’t speak, I might directly mute the sound.}" Additionally, some DHH livestreamers played unrelated background audio, which may also affect hearing audience's understanding, to avoid platform restrictions. DL4 explained, "\textit{We found the platform may categorize our silent livestream as 'vulgar' and ban our account. So now we play music or narrate stories to prevent this.}"

\subsubsection{Challenges in emotional expression and resonance}

In e-commerce livestreaming, hearing livestreamers convey emotions through tone, volume, facial expressions, and gestures, which enhance viewer engagement and purchasing intent \cite{bharadwaj2022new,yu2022just}. However, this form of emotional communication relies heavily on auditory cues, which are absent in SL livestreaming. Instead, DHH livestreamers primarily express emotions through facial expressions and signing styles, which hearing audience may not be familiar with, leading to challenges in emotional connection. Our interviews showed that both DHH livestreamers and hearing audience identified these challenges and mentioned the reasons from their perspectives.

\textbf{Insufficient emotional expression of DHH livestreamers.} Expressing emotions effectively during livestreaming is challenging for DHH livestreamers. Despite two years of experience, DL4 still acknowledged a gap in emotional conveyance compared to experienced livestreamers. DL1 described the reason as a lack of confidence, "\textit{I often feel like I can’t handle it well. Although I can repeat lines like a robot to get the job done, I still feel I need more empathy.}" DL2 also noted the pressure of maintaining enthusiasm despite negative viewer comments. He said, "\textit{Livestreamers need to be mentally strong to maintain a good and positive spirit in livestreaming.}"

\textbf{Misinterpretation of emotions by audience.} Hearing audiences found themselves struggling to interpret and resonate with the emotions expressed in SL livestreaming. They might misinterpret the positive emotions of DHH livestreamers. HV12 misread DHH livestreamers’ enthusiastic signing as aggressive and said, "\textit{They had exaggerated facial expressions, seeming like they were arguing.}" HV8 shared a similar impression: "\textit{They seemed hurried, possibly rushing product introductions.}" Even when hearing audiences correctly identified emotions, they often lacked context to relate and resonate with the emotions. HV2 noted, "\textit{I could sense their excitement but didn’t know why. It felt dull to me.}" HV6 added, "\textit{I can’t relate to their excitement since I don’t understand SL.}"

\section{Virtual co-presenter demo design and evaluation}

\subsection{Virtual co-presenter demo design workshop}

Building on the challenges identified in the previous interviews, we conducted a design workshop to develop a virtual human as a co-presenter for DHH livestreamers and to create a preliminary demo for initial evaluation to guide future design refinements. We recruited six designers with expertise in animation design and user experience design through snowball sampling in local universities. Full demographic details are provided in Appendix A Table \ref{tab:designer}.

The workshop began with a review of SL livestreaming recordings and the communication barriers identified in the interviews. Then, the designers proposed several potential solutions, including captions, a VH, and verbal broadcasts. They discussed and evaluated the advantages, limitations, and feasibility of each based on current technologies. Prior research suggests that captions may distract hearing audience from DHH livestreamers \cite{10.1145/2998181.2998203}, while verbal broadcasts may contribute to ineffective auditory information, as noted in the interviews. Additionally, automatic Chinese SL translation remains underdeveloped \cite{jiang2020survey,10.1145/3544548.3581286}. Based on these considerations, the designers reached a consensus to use VH technology creating a prerecorded virtual co-presenter that leverages the structured nature of SL e-commerce livestreaming, which follows fixed processes and well-prepared scripts.

\begin{figure*}[h!]
    \centering 
    \includegraphics[width=0.8\textwidth]{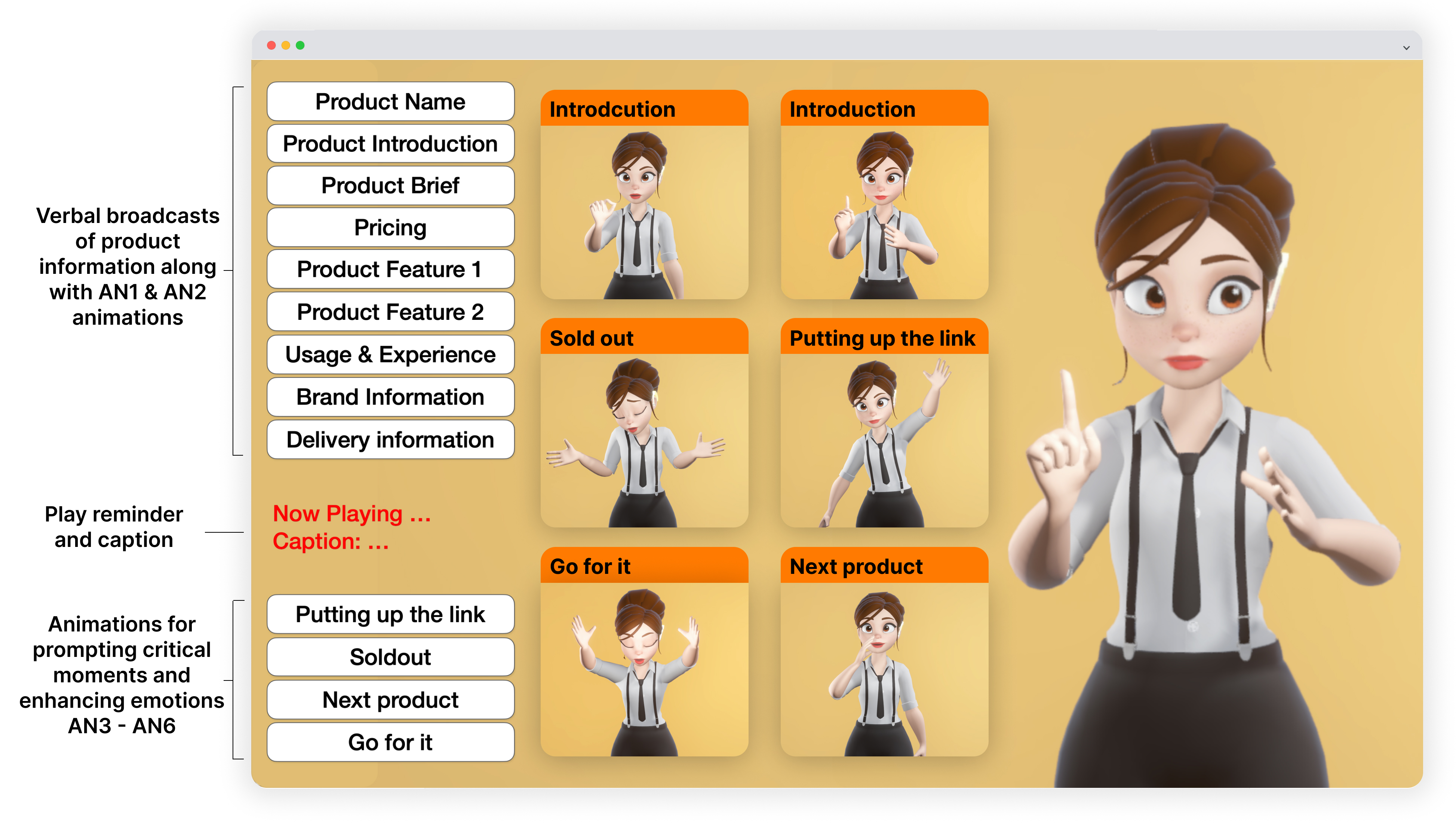}
    \caption{Virtual co-presenter demo.}
    \label{fig:interface}
\end{figure*}

The design process began with selecting the virtual co-presenter’s appearance, opting for a simple, young, human-like cartoon character based on prior research \cite{zhang2017investigating,chen2024avatars}. To help hearing audience better understand DHH livestreamers’ emotions, the designers created six gesture animations (AN1–AN6) shown in Fig. \ref{fig:animations}, imitating gestures from real hearing livestreamers. The selection of gestures was informed by over 30 hours of observed e-commerce livestreaming by hearing livestreamers. AN1 and AN2 were designed to appear randomly or be repeated during the virtual co-presenter verbal broadcasts of product information to improve the anthropomorphism and naturalness of the virtual co-presenter. AN3 to AN6, paired with specific verbal broadcasts, were designed to highlight critical moments and amplify positive emotions. 

After finalizing the gesture animations, the designers developed the virtual co-presenter tool's user interface (Fig. \ref{fig:interface}) for SL livestreaming teams with DHH members. The left panel included buttons triggering verbal broadcasts paired with animations. The upper-left buttons corresponded to nine product information segments provided by DLs, while the lower-left buttons activated AN3–AN6. To enhance accessibility for SL livestreaming teams with DHH members, the designers incorporated a play reminder and red-colored captions to improve visibility and usability.

\subsection{Initial evaluation}

To inform the future development of a full virtual co-presenter system, we conducted an initial evaluation to gather feedback on the preliminary virtual co-presenter demo through semi-structured interviews with the original participants from Section 4. Participants reviewed the virtual co-presenter demo created in the design workshop, providing suggestions on its appearance, functionality, and potential improvements. Hearing audiences watched recordings of virtual co-presenter animations, while SL livestreaming team members evaluated the preliminary virtual co-presenter tool with user interface. All participants from the Section 4 interviews took part in this evaluation. The interviews with DLs lasted between 20 minutes to 40 minutes and the interviews with HVs lasted between 10 minutes to 31 minutes. Data analysis followed the same process as in the previous interview study.

\subsection{Design suggestions for virtual co-presenter demo}

The preliminary virtual co-presenter demo received mostly positive feedback. HV9 noted, "\textit{I think virtual co-presenter will make their livestreaming stand out and be more special.}" HV13 liked the virtual co-presenter demo and said, "\textit{Using virtual co-presenter to help people with language barriers will be great. It will be a trend in the future.}" In addition to positive feedback, both DLs and HVs provided design suggestions regarding the virtual co-presenter's appearance and functionality to improve its role in SL livestreaming.

\subsubsection{Appearance}

Hearing audiences paid more attention to the virtual co-presenter's appearance than SL livestreamers. Most hearing audiences agreed that the virtual co-presenter should resemble a human-like cartoon character. HV3 felt the current design met expectations, describing it as "\textit{an approachable person and not too extravagant.}" HV13 preferred a simple design and stated, "\textit{It doesn’t need too many decorations, just something simple.}" HV1 suggested, "\textit{Its movements and speaking should be human-like, not too mechanical or fake.}" Some hearing audiences also found symbolic or anthropomorphic characters acceptable. HV4 gave an example: "\textit{Many platforms and livestreaming teams have mascots, like the human-like cat of TMall (a Chinese e-commerce platform).}"

Both SL livestreamers and hearing audiences suggested incorporating elements to represent DHH identity. DL1 expressed a preference for the virtual co-presenter to include "\textit{DHH characteristics, such as wearing hearing aids or cochlear implants.}" HV6 also suggested, "\textit{Adding DHH symbols could help hearing audience better recognize their identity if DHH livestreamers are willing to do so.}"

\subsubsection{Functions}

\textbf{Assistance in conveying information.} Almost all hearing audiences expected the prerecorded virtual co-presenter to act as an interpreter to provide real-time SL translation naturally. HV5 emphasized, "\textit{The virtual co-presenter and DHH signers need to be consistent, not only in terms of product information but also in livestreaming processes.}" HV15 suggested that the virtual co-presenter could provide prompts at critical moments to prevent missing key information. HV3 proposed a more interactive role, suggesting, "\textit{For selling food products, the virtual co-presenter could simulate eating and swallowing actions to enhance sensory engagement.}"

\textbf{Assistance in enhancing emotions.} DL1 hoped the virtual co-presenter could help convey "\textit{positive emotions, joy, and a sense of closeness.}" Participants emphasized that the virtual co-presenter’s voice should sound natural, with variations in tone, to avoid a mechanical feel. HV5 believed that "\textit{if the virtual co-presenter expresses emotions, it might encourage purchasing intention.}" HV4 suggested using promotional phrases, such as "\textit{limited quantities available}" or "\textit{today’s lowest price}," to create urgency and boost sales.

\textbf{Controllability and usability.} As primary users of the virtual co-presenter system, DLs stressed the importance of maintaining control. DL3 said, "\textit{The virtual co-presenter should be something we can control.}" This concern also extended to the virtual co-presenter's display position. DL2 preferred that the virtual co-presenter "\textit{not take up too much space or distract audience. It could be placed in a small box in the corner.}" 

Ease of use was another critical factor for DLs. DL2 emphasized, "\textit{We should be able to operate it immediately without the need for extensive training. It would be helpful if it could be triggered directly based on SL.}"

\section{Conclusion and Ongoing Work}

This research investigates communication barriers between DHH livestreamers and hearing audience in sign language e-commerce livestreaming, focusing on challenges in information conveyance and emotional resonance. Through an in-depth analysis of SL livestreaming characteristics, we developed a virtual co-presenter demo via design workshops, aiming to support DHH livestreamers in creating more expressive and interactive content to engage a broader hearing audience. Preliminary evaluations with participants from both communities provided valuable design insights.

Building on these insights, we are refining the virtual co-presenter into a comprehensive system optimized for SL livestreaming team collaboration. With recent advancements in text-to-3D human animation and direct animation generation technologies \cite{azadi2023make,kling2024}, we plan to enhance customization capabilities in future system, allowing DHH livestreamers to flexibly create animated content based on livestreaming scripts. Additionally, we plan to conduct system testing on real livestreaming platforms, iteratively optimizing real-time interaction mechanisms based on feedback from SL livestreaming teams and hearing audiences to assist DHH livestreamers' expressiveness. Our ultimate goal is to establish a design space informed by the needs of both SL livestreaming teams with DHH members and hearing audiences, providing a foundation for broader interactive design between the two communities in e-commerce livestreaming.

\begin{acks}
    We acknowledge the contributions of Yun Wang and Weiwei Zhang in facilitating the workshop and interview. We also appreciate the valuable suggestions from Dr. Hernisa Kacorri.
\end{acks}

\bibliographystyle{ACM-Reference-Format}
\bibliography{ref}

\clearpage
\appendix

\section{Hearing Viewers in Study 1}

\begin{table}[h]
  \caption{Full demographics of hearing viewers.}
  \label{tab:hearingviewer}
  \begin{tabularx}{\textwidth}{c c c p{2.5cm} c c p{7cm}}
    \toprule
    \textbf{ID} &
  \textbf{Gender} &
  \textbf{Age} &
  \multicolumn{1}{c}{\textbf{Platform}} &
  \textbf{\begin{tabular}[c]{@{}c@{}}Frequency\\ (times/month)\end{tabular}} &
  \textbf{\begin{tabular}[c]{@{}c@{}}Duration \\ (minutes)\end{tabular}} &
  \multicolumn{1}{c}{\textbf{Viewing Topics}} \\
    \midrule
    HV1 & F & 29 & Taobao \& Douyin & >8 & 10-30 & Daily Necessities, Food, Cosmetics, Clothes\\
    \hline
    HV2 & M & 23 & Douyin \& Kuaishou & >8 & 10-30 & Daily Necessities, Food, Cosmetics, Electronic Products, Clothes \\
    \hline
    HV3 & M & 19 & Douyin & >8 & 10-30 & Daily Necessities, Food, Cosmetics, Clothes\\
    \hline
    HV4 & F & 26 & Taobao \& Douyin \& Kuaishou & 2-8  & 10-30 & Daily Necessities, Food, Cosmetics, Electronic Products\\
    \hline
    HV5 & M & 23 & Taobao \& Douyin & >8 & 10-30 & Daily Necessities, Food\\
    \hline
    HV6 & F & 28 & Taobao \& Douyin \& Xiaohongshu & >8 & 10-30 & Daily Necessities, Food, Cosmetics\\
    \hline
    HV7 & M & 22 & Douyin \& Kuaishou & >8 & 30-60 & Daily Necessities, Food, Cosmetics, Electronic Products \\
    \hline
    HV8 & M & 20 & Douyin \& Kuaishou & >8 & >60 & Daily Necessities, Food, Maternal and Child Products, Electronic Products \\
    \hline
    HV9 & M & 22 & Taobao \& Douyin \& Xiaohongshu & >8 & >60 & Daily Necessities, Food, Maternal and Child Products, Electronic Products \\
    \hline
    HV10 & F & 33 & Douyin \& Kuaishou & >8 & >60 & Daily Necessities, Food, Cosmetics, Clothes, Maternal and Child Products, Electronic Products  \\
    \hline
    HV11 & M & 28 & Taobao \& Douyin & >8 & 30-60 & Daily Necessities, Food, Cosmetics, Clothes, Electronic Products\\
    \hline
    HV12 & M & 20 & Taobao \& Douyin \& Kuaishou & >8 & 30-60 & Daily Necessities, Food, Cosmetics, Sports Products, Electronic Products, Others\\
    \hline
    HV13 & F & 32 & Taobao \& Douyin & >8 & 30-60 & Daily Necessities, Food, Cosmetics,  Electronic Products \\
    \hline
    HV14 & M & 25 & Taobao \& Douyin & >8 & 30-60 & Daily Necessities, Food, Cosmetics,  Clothes \\
    \hline
    HV15 & F & 30 & Taobao & >8 & 10-30 & N/A \\
    
    \bottomrule
  \end{tabularx}
\aptLtoX[graphic=no,type=html]{}{\begin{minipage}{\textwidth}\smallskip}
    \small "Platform" describes the applications that hearing viewers use to watch livestreaming. "Frequency (times/month)" describes the average frequency of hearing viewers watching livestreaming in their daily life. "Duration (minutes)" describes the average time of hearing viewers watching livestreaming each time in their daily life. "Viewing Topics" describes the categories that hearing viewers watch livestreaming for in their daily life.
\aptLtoX[graphic=no,type=html]{}{\end{minipage}}
\end{table}

\begin{table}[h]
\caption{Full demographics of designers.}
\label{tab:designer}
\begin{tabular}{ccccc}
\hline
\textbf{Designer} & \textbf{Gender} & \textbf{Age} & \textbf{Expertise}                & \textbf{\begin{tabular}[c]{@{}c@{}}Reported watching time of\\ hearing e-commerce livestreaming\end{tabular}} \\ \hline
D1                & M               & 27           & 6 years in Animation Design       & 4 hours                                                                                                      \\
D2                & F               & 24           & 3 years in User Experience Design & 5$\sim$6 hours                                                                                               \\
D3                & M               & 21           & 3 years in Animation Design       & 4$\sim$5 hours                                                                                               \\
D4                & F               & 35           & 10 years in User Interface Design & more than 6 hours                                                                                            \\
D5                & F               & 22           & 4 years in User Experience Design & about 5 hours                                                                                                \\
D6                & F               & 21           & 3 years in Animation Design       & more than 5 hours                                                                                           \\ \hline
\end{tabular}
\end{table}

\clearpage
\section{Virtual co-presenter's animations and corresponding quotes}

\begin{figure}[!h]
    \centering
    \includegraphics[width=0.75\textwidth]{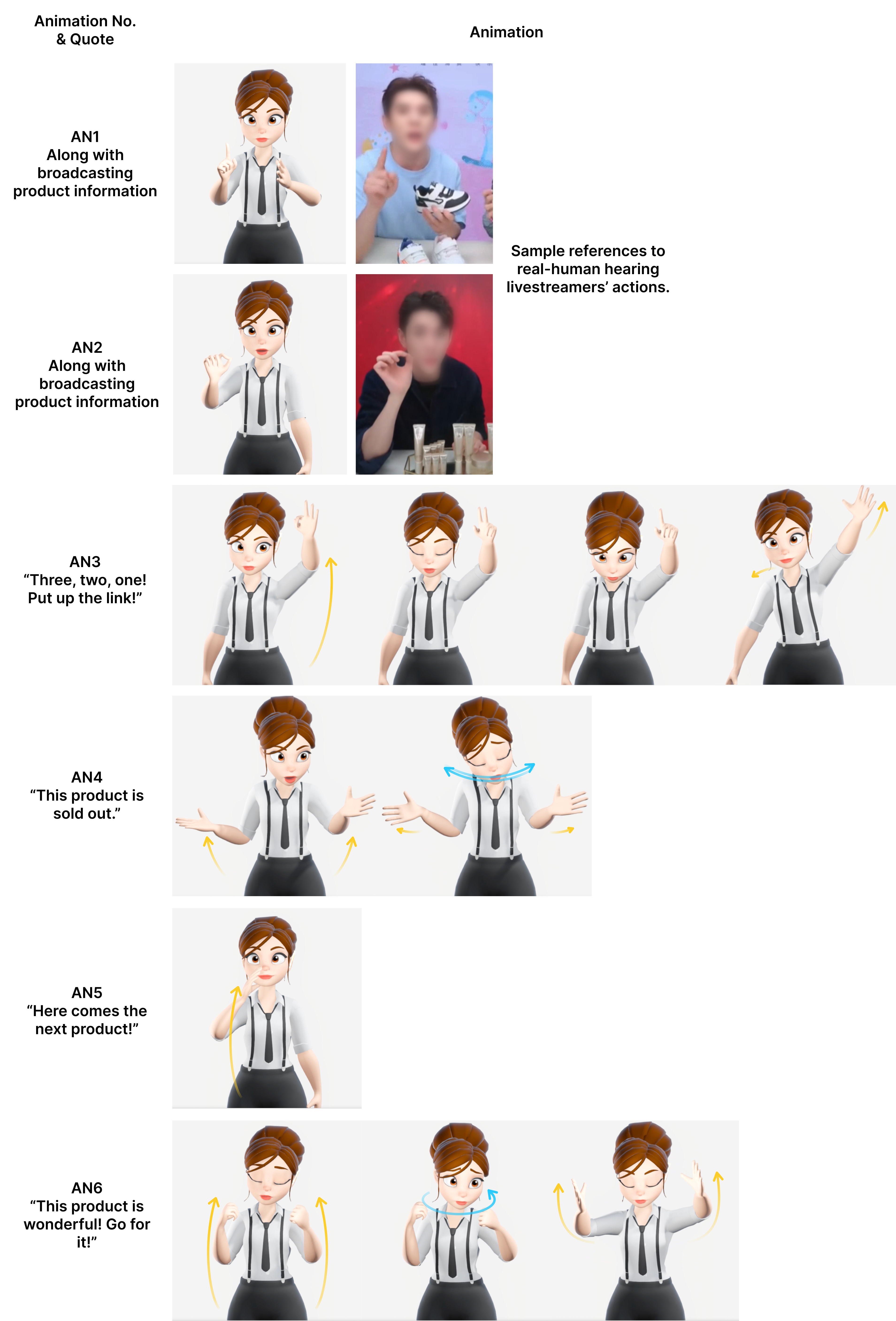}
    \caption{Animations and corresponding quotes (AN1-AN6).}
    \label{fig:animations}
\end{figure}

\end{document}